\documentclass[preprint,prl,twocolumn,10pt]{revtex4}%
\usepackage{amssymb}
\usepackage{amsfonts}
\usepackage{amsmath}
\usepackage{graphicx}%
\providecommand{\U}[1]{\protect\rule{.1in}{.1in}}

\begin{document}
\preprint{ }
\title{Renormalized Onsager functions and merging of vortex clusters}
\author{Franco Flandoli}
\affiliation{Scuola Normale Superiore of Pisa}

\begin{abstract}
In this letter we numerically investigate the merging mechanism between two
clusters of point vortices. We introduce a concept of renormalized Onsager
function, an elaboration of the solutions of the mean field equation, and use
it to understand the shape of the single cluster observed as a result of the
merging process. We finally discuss the potential implications for the inverse
cascade 2D turbulence.

\end{abstract}
\maketitle

\textit{Introduction} - Stationary inverse cascade in 2D\ fluids with small
scale activation and large scale friction is observed in experiments and
quantified quite well by scaling laws, see the reviews \cite{Tabeling},
\cite{Boffetta}. Dimensional analysis for the average square velocity
increments $u_{r}^{2}:=\left\langle \left\vert u\left(  x+r\right)  -u\left(
x\right)  \right\vert ^{2}\right\rangle $ assumes that $u_{r}^{2}$ depends
only on $r$ and the energy flux $\epsilon$ and has a scaling law $u_{r}%
^{2}=C\epsilon^{\alpha}r^{\beta}$. Equating $\left[  L\right]  ^{2}/\left[
T\right]  ^{2}$ to $\left[  L\right]  ^{2\alpha}/\left[  T\right]  ^{3\alpha
}\cdot\left[  L\right]  ^{\beta}$ one immediately gets $\alpha=\frac{2}{3}$
and $\beta=\frac{2}{3}$, namely $u_{r}=C\epsilon^{1/3}r^{1/3}$. This simple
argument gives a result which was never contradicted by experiments. But it
does not explain the mechanisms of the inverse cascade, it is based on
assumptions which are not directly verifiable, and it gives the same result in
3D, where experiments reveal important deviations.

Vortex structures are certainly involved in the inverse cascade. One of the
mechanisms which could be relevant is the aggregation of vortices in larger
and larger clusters. This note aims to contribute to the understanding of one
fragment of this complex process, namely the merging of two clusters of
vortices into a larger one. Starting from the mean field equation of Onsager
theory, we introduce the concept of renormalized Onsager function. It is a
family of functions parametrized by a real number $\beta$ (including the
negative values promoted by Onsager) which correspond to unitary variance
configurations and, properly rescaled, covers the class of all solutions of
the mean field equation. We observe that the shapes emerging in very short
time (between an half and one turnover time)\ from the merging of two roughly
similar and close clusters of equal sign is very close to a renormalized
Onsager function; more precisely, this is true in a class of numerical
experiments, while in others there is a systematic deviation that requires
further study but is addressed here by a first rough correction. Finally, we
speculate how the results found here could be the starting point of a theory
of inverse cascade, yelding also $u_{r}=C\epsilon^{1/3}r^{1/3}$ (up to
logarithmic corrections).

\textit{Renormalized Onsager functions - }Consider $N$ point vortices
$X_{1},...,X_{N}$ in the plane, each one with circulation $\Gamma$; the
vorticity field is $\frac{\Gamma}{N}\sum_{i=1}^{N}\delta\left(  x-X_{i}%
\right)  $. Kinetic energy is infinite but modulated by the finite quantity
(which corresponds to interaction energy) $\mathcal{H}:=-\sum
_{\substack{i,j=1\\i\neq j}}^{N}\frac{\Gamma^{2}}{N^{2}}\log\left\vert
X_{i}-X_{j}\right\vert $, invariant for the vortex dynamics $\frac{dX_{i}}%
{dt}=\sum_{j\neq i}\frac{\Gamma}{2\pi}\frac{\left(  X_{i}-X_{j}\right)
^{\perp}}{\left\vert X_{i}-X_{j}\right\vert ^{2}}$. Two other relevant
invariants are the center of mass $\mathcal{M}=\frac{1}{N}\sum_{i=1}^{N}X_{i}$
and the variance $\mathcal{V}=\frac{1}{N}\sum_{i=1}^{N}\left\vert
X_{i}-\mathcal{M}\right\vert ^{2}$, (related to the moment of inertia). Given
two numbers $e,\sigma$, $\sigma\geq0$ and a point $m\in\mathbb{R}^{2}$,
consider the microcanonical measure formally defined by $\mu_{N}^{e,\sigma
,m}=\delta\left(  \mathcal{H=}\Gamma^{2}e,\mathcal{M}=m,\mathcal{V}=\sigma
^{2}\right)  $. Onsager theory \cite{Onsager}, \cite{EySree}, complemented by
a more explicit formulation of the mean field equation by Montgomery and Joice
\cite{Montgomery} and by various rigorous results (see for instance
\cite{Caglioti}, \cite{Eyik}, \cite{Lions}) claims that
\[
\int_{\mathbb{R}^{2N}}d\mu_{N}^{e,\sigma,m}\left\vert \frac{\Gamma}{N}%
\sum_{i=1}^{N}\varphi\left(  x_{i}\right)  -\Gamma\int_{\mathbb{R}^{2}}%
\varphi\left(  x\right)  \rho_{\alpha,\beta}\left(  x-m\right)  dx\right\vert
^{2}%
\]
converges to zero for every smooth compact support test function $\varphi$ on
$\mathbb{R}^{2}$. Here the pair $\left(  \alpha,\beta\right)  $ (with
$\alpha\geq0$) is uniquely prescribed by $\left(  e,\sigma\right)  $ and
$\rho_{\alpha,\beta}\left(  x\right)  $ is a probability density function
given by $\rho_{\alpha,\beta}\left(  x\right)  =Z_{\alpha,\beta}^{-1}%
e^{-\beta\phi_{\alpha,\beta}\left(  x\right)  -\alpha\left\vert x\right\vert
^{2}}$, $Z_{\alpha,\beta}=\int e^{-\beta\phi_{\alpha,\beta}\left(  x\right)
-\alpha\left\vert x\right\vert ^{2}}dx$, where $\phi_{\alpha,\beta}\left(
x\right)  $ is the solution of the mean field equation%
\[
\Delta\phi_{\alpha,\beta}\left(  x\right)  =-Z_{\alpha,\beta}^{-1}%
e^{-\beta\phi_{\alpha,\beta}\left(  x\right)  -\alpha\left\vert x\right\vert
^{2}}.
\]
Uniqueness is true under the condition that the velocity $\nabla^{\perp}%
\phi_{\alpha,\beta}$ vanishes at infinity and that $\phi_{\alpha,\beta}$ is
directly linked to the interaction energy $\mathcal{H}$ above, which reduces
to the condition $\phi_{\alpha,\beta}\left(  0\right)  =-\frac{1}{2\pi}%
\int\log\left\vert x\right\vert \rho_{\alpha,\beta}\left(  x\right)  dx$. A
dynamical theory of convergence to equilibrium is however missing
\cite{EySree}. There are initial configurations, like those corresponding to
rotation invariant profiles, having a time of convergence to equilibrium that
is essentially infinite. However, other initial configurations have a much
shorter relaxation time, if we accept some degree of approximation;\ this is
what we want to describe with the following numerical experiments.

Solutions of the mean field equation are rotationally invariant; with little
abuse of notation we shall write $\rho_{\alpha,\beta}\left(  r\right)  $,
$\phi_{\alpha,\beta}\left(  r\right)  $ as functions of the distance to the
center of mass. They have a special scaling property in $\alpha$: let us call
\textit{canonical} case the equation with $\alpha=1$, whose solutions will be
denoted by $\rho_{\beta}\left(  x\right)  $, $\phi_{\beta}\left(  x\right)  $.
In this case we impose $\phi_{\beta}\left(  0\right)  =0$ and $\nabla
\phi_{\beta}\left(  0\right)  =0$, convenient for numerical purposes (the
condition $\nabla\phi_{\beta}\left(  0\right)  =0$ is motivated by radial
symmetry and differentiability at the origin). Then a simple computation shows
that
\[
\phi_{\alpha,\beta}\left(  x\right)  =\phi_{\beta}\left(  \sqrt{\alpha
}x\right)  +C_{\alpha,\beta},\qquad\rho_{\alpha,\beta}\left(  x\right)
=\alpha\rho_{\beta}\left(  \sqrt{\alpha}x\right)
\]%
\[
Z_{\alpha,\beta}=\alpha^{-1}e^{-\beta C_{\alpha,\beta}}Z_{\beta},\qquad
C_{\alpha,\beta}=\frac{1}{2\pi}\int\lg\frac{\sqrt{\alpha}}{\left\vert
x\right\vert }\rho_{\beta}\left(  x\right)  dx.
\]
Thus it is sufficient to know the shapes $\rho_{\beta}\left(  x\right)  $,
$\phi_{\beta}\left(  x\right)  $ and rescale them as above.

However, comparing $\rho_{\beta}\left(  x\right)  $ for different values of
$\beta$ is not so useful. In examples, we are given an initial family of
vortex points with a value of $\left(  e,\sigma,m\right)  $. We should find a
pair $\left(  \alpha,\beta\right)  $ such that
\begin{align*}
\int\rho_{\alpha,\beta}\left(  x\right)  \phi_{\alpha,\beta}\left(  x\right)
dx &  =e\\
\int\left\vert x\right\vert ^{2}\rho_{\alpha,\beta}\left(  x\right)  dx &
=\sigma^{2}%
\end{align*}
(one can see that, with the prescriptions above, $\Gamma^{2}\int\rho
_{\alpha,\beta}\left(  x-m\right)  \phi_{\alpha,\beta}\left(  x-m\right)  dx$
is the continuum analog of $\mathcal{H}$, and obviously $\int\left\vert
x-m\right\vert ^{2}\rho_{\alpha,\beta}\left(  x-m\right)  dx$ is the continuum
analog of $\mathcal{V}$). Although theoretically these equations are on the
same ground, at a practical level the second one, $\int\left\vert x\right\vert
^{2}\rho_{\alpha,\beta}\left(  x\right)  dx=\sigma^{2}$, imposes a quite
strict and stable constraint, while the first one is relatively poor, because
the typical values of $E$ are very small and with imperceptible, logarithmic
variations for moderate changes of the initial configuration. Said
differently, the value of $\sigma$ is very relevant in practice, while the
value of $e$ is less easy to use in numerical experiments. The second equation
gives us $\alpha=\sigma^{-2}\int\left\vert x\right\vert ^{2}\rho_{\beta
}\left(  x\right)  dx$. Let us set $\sigma_{\beta}^{2}:=\int\left\vert
x\right\vert ^{2}\rho_{\beta}\left(  x\right)  dx$. Thus, given an initial
configuration with a value of $\sigma$, we may parametrize Onsager shapes
directy by $\left(  \sigma,\beta\right)  $:%
\[
\rho_{\alpha,\beta}\left(  x\right)  =\sigma^{-2}\widetilde{\rho}_{\beta
}\left(  \sigma^{-1}x\right)  \text{,\qquad\ }\widetilde{\rho}_{\beta}\left(
x\right)  :=\sigma_{\beta}^{2}\rho_{\beta}\left(  \sigma_{\beta}x\right)  .
\]
Notice that $\int\left\vert x\right\vert ^{2}\widetilde{\rho}_{\beta}\left(
x\right)  dx=1$. We call $\widetilde{\rho}_{\beta}$ \textit{renormalized
Onsager functions}. Comparing $\widetilde{\rho}_{\beta}\left(  x\right)  $ is
the starting step to understand possible emerging shapes. In Figure 1 we
compare the cases $\beta=-15,0,30$ by plotting $f_{R}\left(  r\right)  =2\pi
r\rho_{\alpha,\beta}\left(  r\right)  $, the probability density function of
the distance from the center of mass, and $F_{R}\left(  r\right)  $, the
corresponding cumulative distribution function (cdf), which will be used below
in numerical experiments.%

\begin{figure}[ptb]%
\centering
\includegraphics[
height=2.8163in,
width=2.8163in
]%
{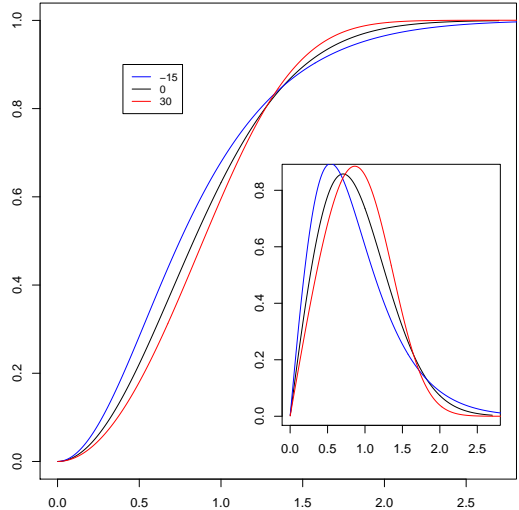}%
\caption{Cdf of the distance from the center of mass for three examples of
renormalized Onsager shapes, $\beta=-15,0,30$ (pdf in the small figure).}%
\end{figure}

The value of $\sigma$ determines the typical distance of points from the
center.\ The parameter $\beta$ modulates only a little bit the shape. For
$\beta=0$ the density $\widetilde{\rho}_{0}\left(  x\right)  =\sigma^{-2}%
Z_{0}^{-1}e^{-\left\vert \sigma^{-1}x\right\vert ^{2}}$ is Gaussian, the
maximum entropy distribution among those with a given variance, here equal to
one. For $\beta<0$ the unitary variance constraint is maintained by means of
more points close to the center of mass and more points far from it. For
$\beta>0$ points tend to stay closer to unitary distance from the center of
mass with respect to the Gaussian. In bounded domains the constraint of
constant variance cannot be imposed and the role of $\beta$ is more striking.
Here in full space it plays a role of correction over the shape imposed by
unitary variance.

Simulations of\ canonical Onsager mean field equation, for a given $\beta$,
are made using the equation for the radial component%
\[
\frac{1}{r}\phi_{\beta}^{\prime}\left(  r\right)  +\phi_{\beta}^{\prime\prime
}\left(  r\right)  =-Z_{\beta}^{-1}e^{-\beta\phi_{\beta}\left(  r\right)
-r^{2}}%
\]
with $\phi_{\beta}\left(  0\right)  =\phi_{\beta}^{\prime}\left(  0\right)
=0$, finding the right value of $Z_{\beta}=2\pi\int_{0}^{\infty}e^{-\beta
\phi_{\beta}\left(  r\right)  -r^{2}}rdr$ by iteration until the value is
sufficiently stabilized. The nonphisical (but numerically useful) condition
$\phi_{\beta}\left(  0\right)  =0$ is then removed by the constant
$C_{\alpha,\beta}$ above. The cumulative distribution function of the radius
is computed as $F_{R}\left(  r\right)  =-2\pi\left(  \sqrt{\alpha}r\right)
\phi_{\beta}^{\prime}\left(  \sqrt{\alpha}r\right)  $.

\textit{Merging of two clusters} - In this section we investigate numerically
the merging process between two clusters of point vortices, all with the same
circulation, that we normalize so that the pair of clusters is globally a
probability measure. The continuum limit, at time zero, is assumed to have the
form%
\[
\omega_{0}\left(  x\right)  =\frac{1}{2}\rho_{1}\left(  x-\frac{d}{2}%
e_{1}\right)  +\frac{1}{2}\rho_{2}\left(  x+\frac{d}{2}e_{1}\right)
\]
$e_{1}=\left(  1,0\right)  $, where $\rho_{1}$ and $\rho_{2}$ are probability
denstities, hence $\omega_{0}$ is as well. The pdf $\rho_{1}$ and $\rho_{2}$
may be different and, up to small variations, will have unitary variance. By
rescaling space and time, the understanding of this model case is
representative of any size and any circulation. We approximate $\omega
_{0}\left(  x\right)  $ by two clusters of independent points
\[
\frac{1}{2N}\sum_{i=1}^{N}\delta\left(  x-X_{i}\right)  +\frac{1}{2N}%
\sum_{j=1}^{N}\delta\left(  x-Y_{j}\right)
\]
with $X_{i}$ (resp. $Y_{j}$) distributed as $\rho_{1}\left(  x-\frac{d}%
{2}e_{1}\right)  $ (resp. $\rho_{2}\left(  x+\frac{d}{2}e_{1}\right)  $). When
the distance $d$ is large compared to the size $r$, the two structures rotate
around their center of mass (and each one around its own center of mass) like
two point vortices, just experiencing some degree of deformation of the
circular structure in a roughly ellipsoidal one; this vortex-patch dynamics,
approximating point vortex one, has been well understood by
\cite{MarPulvpatches}.

%

\begin{figure}[ptb]%
\centering
\includegraphics[
height=2.8163in,
width=2.8163in
]%
{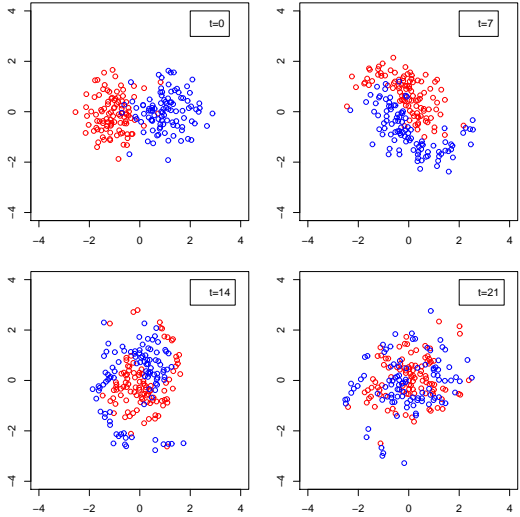}%
\caption{Merging process of two unitary variance Gaussian clusters at distance
$d=2$. The four pictures show the initial configuration and three instances
during the first turnover.}%
\end{figure}

On the the contrary, when $d$ is small, typically of the order of $2-3$ times
the "radius" of the structures, the two clusters start immediately a merging
process which evolves into a new larger cluster. Based on rigorous convergence
results of point vortices to Euler equations \cite{Marchioro}, \cite{Shoquet},
we know that
\begin{equation}
\frac{1}{2N}\sum_{k=1}^{2N}\delta\left(  x-Z_{k}\left(  t\right)  \right)
\sim\rho\left(  t,x\right)  \label{new structure}%
\end{equation}
namely $Z_{k}\left(  t\right)  $ are distributed as the probability density
$\rho\left(  t,x\right)  $ which solves Euler equations with initial condition
$\omega_{0}\left(  x\right)  $. The aim of our investigation is to identify an
approximate shape for $\rho\left(  t,x\right)  $, based on Onsager theory,
valid for relatively small $t$ (around one turnover time). The initial
configuration $\left(  X_{i},Y_{j},i,j=1,...,N\right)  $ lives on the surface
$\mathcal{H}=e$, $\mathcal{V}=\sigma^{2}$, but it is "anomalous" with respect
to the typical configurations described by $\rho_{\alpha,\beta}\left(
x\right)  $ with $\left(  \alpha,\beta\right)  $ corresponding to $\left(
e,\sigma\right)  $. Statistical mechanics predicts convergence to
$\rho_{\alpha,\beta}\left(  x\right)  $. As remarked above, in principle there
are initial configurations which take too much time for convergence; what we
observe numerically is a substantial approach to $\rho_{\alpha,\beta}\left(
x\right)  $ in the time of one turnover or less.%

\begin{figure}[ptb]%
\centering
\includegraphics[
height=2.8163in,
width=2.8163in
]%
{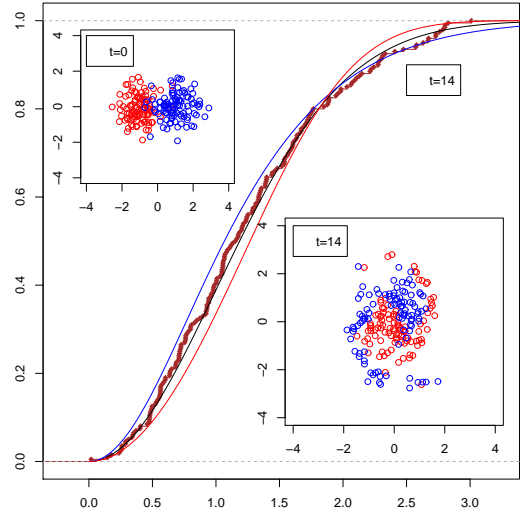}%
\caption{Comparison between empirical cdf (of the distance from the center of
mass) and Onsager renormalized functions for $\beta=-15,0.30$, for the data of
Figure 2. }%
\end{figure}
%

\begin{figure}[ptb]%
\centering
\includegraphics[
height=2.8163in,
width=2.8163in
]%
{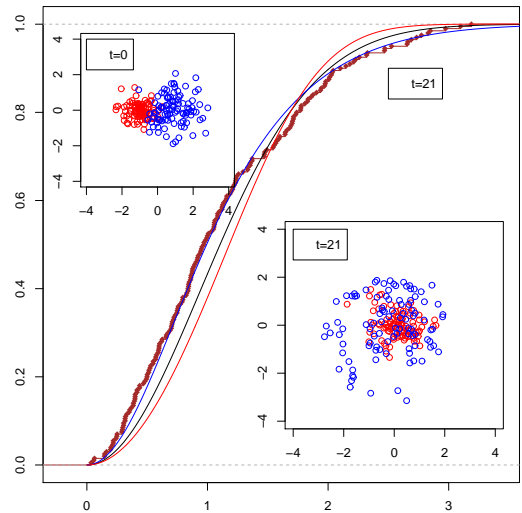}%
\caption{An example with moderately different size of the initial clusters.}%
\end{figure}

We report several experiments. The first one is the case of two Gaussian
clusters of unitary variance at distance $d=2$. Turnover time is of the order
of 20 sec (if space is measured in meters). The empitical cdf (ecdf), already
between one-half and one turnover time, is very close to the class of Onsager
functions, precisely to the Gaussian shape itself, $\beta=0$. In Figure 2 we
show vortex configurations at subsequent instants in the first turnover period
of time, proving convergence towards a shape substantially invariant by
rotations. In Figure 3 we show the ecdf even before the first turnover time
(later on it is substantially the same) superimposed to Onsager functions;\ we
introduce the display, used also below, of the ecdf and, in small boxes the
configurations at initial and final time.

This picture is partially stable under certain perturbations. In Figure 4 we
show the case of two unequal initial Gaussian clusters. Here we clearly
observe the choice, by the system, of negative $\beta$. In general negative
$\beta$ arise when, in the initial configuration, there is a remarkable
quantity of vortex points more distant from the center of mass than the bulk
of points. Under such conditions the system has a tendency to develop wings,
namely to loose the boudary points through filamentary structures; while the
bulk concentrates more, to compensate the distant points (due to conservation
of variance). As remarked above when we discussed how renormalized Onsager
functions change with $\beta$, this behavior corresponds to negative $\beta$.%

\begin{figure}[ptb]%
\centering
\includegraphics[
height=2.8163in,
width=2.8163in
]%
{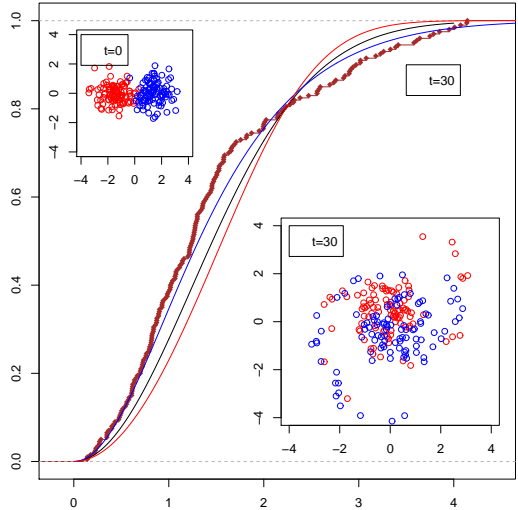}%
\caption{Two unitary Gaussian clusters at distance $d=3$. Deviation from
renormalized Onsager functions is systematic. }%
\end{figure}
%

\begin{figure}[ptb]%
\centering
\includegraphics[
height=2.8163in,
width=2.8163in
]%
{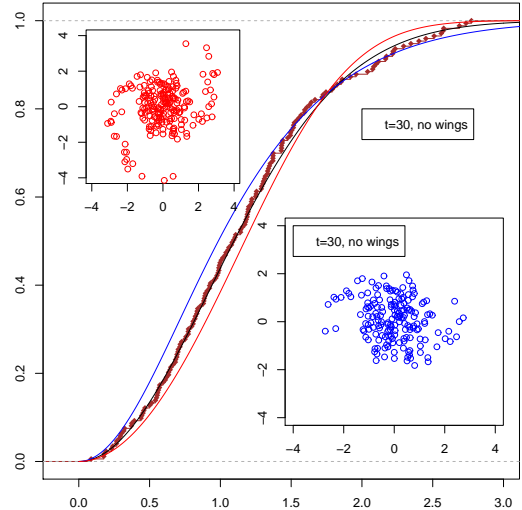}%
\caption{Elimination of wings in the example of Figure 5. The original shape
at time $t=30$ is in the upper box. The lower box and the ecdf correspond to
the case without wings.}%
\end{figure}

\textit{Deviation from renormalized Onsager functions} - When the dispersion
of the initial configuration, in the sense just described above, is too large,
the ecdf is vaguely similar to Onsager functions with $\beta<0$ but it shows
also a systematic deviation. We illustrate this fact in Figure 5 with the case
of two equal unitary Gaussian clusters at distance $d=3$ (instead of $d=2$).
The phenomenology is similar to the one described above: wings of dispersed
points and a strong kernel to compensate for the constant variance. But the
shape is not Onsager anymore. We have not discover yet a variation of Onsager
theory which may incorporate this case. However, a simple argument restores
some fact. If we eliminate the extreme parts of the wings we observe again a
good level of coincidence with renormalized Onsager functions. We show in
Figure 6 the result after the elimination of the tails, in the case of the
data of Figure 5.

\textit{Potential relevance for inverse cascade }- Aggregation of vortex
structures from smaller to larger ones is a well known phenomenon, clearly
related to a cascade of energy from smaller to larger scales. What we may
speculate after the observations of the previous sections is that a turbulent
fluid may be composed, up to a disordered low-intensity component (which
presumably includes the extremal parts of the wings formed during stretching
processes), of localized intense vortex structures having approximatively a
shape invariant by rotation and with radial distribution approximatively equal
to a renormalized Onsager shape, rescaled by the standard deviation of the
structure. These kind of structures are self-consistent, in the sense that two
smaller ones merge into a larger one. The parameter $\beta$ may vary and
sometimes the Onsager shape can be attributed only to the bulk of the
structure, dispersing the wings into the disordered background.

Assume the fluid is maintained in a stationary regime by injection of vortex
structures at very small scale and dissipation by friction. Around every scale
$r_{0}$ we may assume to observe structures approximatively of that size,
namely of the form $r_{0}^{-2}\widetilde{\rho}_{\beta_{0}}\left(  r_{0}%
^{-1}x\right)  $ where $\widetilde{\rho}_{\beta_{0}}$ is a renormalized
Onsager function. When two such structures (with the same sign of circulation)
are sufficiently close, they will merge into a new structure of the form
$r_{1}^{-2}\widetilde{\rho}_{\beta_{1}}\left(  r_{1}^{-1}x\right)  $ with
$r_{1}>r_{0}$ and some $\beta_{1}$ (possibly up to elimination of extreme
parts of wings). Admitting the unjustified simplicity of the next argument,
let us assume we have an injection scale $r_{injection}$ and, instead of a
continuum of scales, only discrete scales
\[
r_{injection}=r_{l_{inj}}<...<r_{l+1}<r_{l}<...<r_{0}%
\]
and that only merging events occur between structures of the same scale
$r_{l+1}$ (for some $l$) producing structures of scale $r_{l}$; and finally
that there is a characteristic circulation $\Gamma_{l}$ associated to scale
$l$. The vorticity field has the form (up to a low-intensity disordered
backgound)%
\begin{equation}
\omega\left(  x\right)  =\sum_{l}\Gamma_{l}\sum_{i\in\Lambda_{l}}r_{l}%
^{-2}\widetilde{\rho}_{\beta_{i}}\left(  r_{l}^{-1}\left(  x-x_{i}^{0}\right)
\right)  \label{self similar}%
\end{equation}
where $\Lambda_{l}$ indexes the set of structures of level $l$. Typical
velocity at scale $l$ is $u_{l}=\frac{\Gamma_{l}}{r_{l}}$ (from Biot-Savart
relation $u\left(  x\right)  =\frac{1}{2\pi}\int\frac{\left(  x-y\right)
^{\perp}}{\left\vert x-y\right\vert ^{2}}\omega\left(  y\right)  dy$). If we
discover a relation between $\Gamma_{l}$ and $r_{l}$, we find a formula for
$u_{l}$ as a function of $r_{l}$, to compare with the scaling law
$u_{r}=C\epsilon^{1/3}r^{1/3}$.

Here $\epsilon$ is the energy per unit of space-time injected at scale
$r_{l_{inj}}$ through the creation of the smaller vortex blobs. The system
behaves like a stationary linear queuing network, with the same energy flux at
each level $l$, and the rule $\epsilon=\lambda_{l}\cdot\epsilon_{l}$, where
$\epsilon_{l}$ is the kinetic energy of one structure of level $l$, while
$\lambda_{l}$ is the average number of "events" at level $l$ in unit of
space-time (either we choose to consider events the new arrivals, or we choose
the departures, it is equivalent). Using the formula $\int\int\log\left\vert
x-y\right\vert \Gamma_{l}\rho_{l}\left(  x\right)  \Gamma_{l}\rho_{l}\left(
y\right)  dxdy$, where we have abbreviated $\rho_{l}\left(  x\right)
=r_{l}^{-2}\widetilde{\rho}_{\beta_{i}}\left(  r_{l}^{-1}\left(  x-x_{i}%
^{0}\right)  \right)  $, we find $\epsilon_{l}\sim\Gamma_{l}^{2}\log\frac
{1}{r_{l}}$. Hence $\epsilon\sim\lambda_{l}\cdot\Gamma_{l}^{2}\log\frac
{1}{r_{l}}$. In queuing theory $\lambda_{l}$ is the throughput, or average
arrival/departure rate. A version of Little's law states that $\lambda
_{l}=\frac{n_{l}}{\tau_{l}}$ where $n_{l}$ is the average number of structures
involved in potential merging events and $\tau_{l}$ is the merging time. By
equilibrium considerations ($n_{l}r_{l}^{2}\sim$ area occupied by merging
structures) it is reasonable to assume that $n_{l}\sim\frac{C}{r_{l}^{2}}$
(this detail is more intricate than others and requires deeper investigation).
The merging time at scale $l$ for structures with circulation $\Gamma_{l}$ is
of the order $\tau_{l}\sim\frac{r_{l}^{2}}{\Gamma_{l}}$, by a\ simple
computation based on the rescaling $\omega\left(  t,x\right)  :=\frac
{r_{l}^{2}}{2\Gamma_{l}}\omega_{r_{l}}\left(  \frac{r_{l}^{2}}{\Gamma_{l}%
}t,r_{l}x\right)  $; but essential is to assume that the merging time at
unitary scale is unitary (up to a constant), fact that we observed in the
numerical simulations above.

Collecting these facts we have $\epsilon=\frac{n_{l}}{\tau_{l}}\epsilon
_{l}\sim C\frac{\Gamma_{l}^{3}}{r_{l}^{4}}\log\frac{1}{r_{l}}$. It follows%
\[
\Gamma_{l}\sim C\epsilon^{1/3}r_{l}^{4/3}/\log^{1/3}\frac{1}{r_{l}}%
\]
hence $u_{l}\sim C\epsilon^{1/3}r_{l}^{1/3}$ (up to logarithmic corrections in
$r_{l}$) which gives the correct scaling law (logarithmic corrections have
been invoked in the literature \cite{Boffetta}, but experiments do not clarify
this issue, due to few scales).

Turbulence is a non-equilibrium time-stationary system. The previous picture
restores a very weak form of local equilibrium. In the classical form of local
equilibrium, arbitrarily small macroscopic portions of the medium go to
equilibrium in arbitrary short time. Here, convergence to equilibrium holds
only for small but well defined portions of fluid and requires a
\textit{macroscopic time}, which does not go to zero with the size (it is an
obvious consequence of the fact that particles, the point vortices, move at
speed comparable to the macroscopic time). However this time is short,
$\tau_{l}\sim\frac{r_{l}^{2}}{\Gamma_{l}}\sim\epsilon^{-1/3}r_{l}^{2/3}%
\log^{1/3}\frac{1}{r_{l}}$ as discussed above, for vorticity configurations
made of two close small vortex blobs.

As a final remark, from this picture emerges an approximate self-similar
picture of the form (\ref{self similar}) with suitable scaling laws of the
parameters. We do not claim here that there is full self-similarity, this
issue requires closer investigation, but it is not unreasonable. This
structure, in the limit $l_{inj}\rightarrow\infty$, could be useful to
investigate more advanced properties like the SLE\ structure of level lines
\cite{Bernard}; the Poissonian structure emerging from the present
description, by analogy with critical percolation, seems to be in favour of
the conjecture SLE(6).

\end{document}